\documentclass[twocolumn, trackchanges]{aastex61}

\usepackage{multirow}
\usepackage{asmath}
\usepackage{graphicx}

\newcommand{\myemail}{m.cornachione@utah.edu}
\newcommand{\Romannumeral}[1]{\uppercase\expandafter{\romannumeral #1\relax}}

\shorttitle{Morphology of $z\approx2.5$ Lensed Ly$\alpha$ Emitters}
\shortauthors{Cornachione et al.}

\begin{document}

\title{The BOSS Emission-Line Lens Survey V. Morphology and Substructure of Lensed Lyman-$\alpha$ Emitters at redshift $z\approx2.5$ in the BELLS GALLERY}

\author{Matthew A. Cornachione}
\affiliation{Department of Physics and Astronomy, University of Utah, 115 South 1400 East,
    Salt Lake City, UT 84112, USA}
\email{\myemail, bolton@noao.edu}

\author{Adam Bolton}
\affiliation{National Optical Astronomy Observatory, 950 N. Cherry Ave., Tucson, AZ 85719, USA}

\author{Yiping Shu}
\affiliation{National Astronomical Observatories, Chinese Academy of Sciences, A20 Datun Rd., Chaoyang District, Beijing 100012, China}
\affiliation{Purple Mountain Observatory, Chinese Academy of Sciences, 2 West Beijing Road, Nanjing 210008, China}

\author{Zheng Zheng}
\affiliation{Department of Physics and Astronomy, University of Utah, 115 South 1400 East,
    Salt Lake City, UT 84112, USA}

\author{Antonio D. Montero-Dorta}
\affiliation{Departamento de F{\'i}sica Matem{\'a}tica, Instituto de F{\'i}sica, Universidade de S{\~a}o Paulo, Rua do Mat{\~a}o 1371, CEP 05508-090, S{\~a}o Paulo, Brazil}
\affiliation{Instituto de Astrof{\'i}sica de Andaluc{\'i}a (CSIC), Glorieta de la Astronom{\'i}a, E-18080 Granada, Spain}
\affiliation{Department of Physics and Astronomy, University of Utah, 115 South 1400 East, Salt Lake City, UT 84112, USA}
    
\author{Joel R. Brownstein}
\affiliation{Department of Physics and Astronomy, University of Utah, 115 South 1400 East,
    Salt Lake City, UT 84112, USA}

\author{Masamune Oguri}
\affiliation{Research Center for the Early Universe, University of Tokyo, 7-3-1 Hongo, Bunkyo-ku, Tokyo 113-0033, Japan}
\affiliation{Department of Physics, University of Tokyo, 7-3-1 Hongo, Bunkyo-ku, Tokyo 113-0033, Japan}
\affiliation{Kavli Institute for the Physics and Mathematics of the Universe (Kavli IPMU, WPI), University of Tokyo, Chiba 277-8583, Japan}

\author{Christopher S. Kochanek}
\affiliation{Department of Astronomy $\&$ Center for Cosmology and Astroparticle Physics, Ohio State University, Columbus, OH 43210, USA}

\author{Shude Mao}
\affiliation{Physics Department and Tsinghua Centre for Astrophysics, Tsinghua University, Beijing 100084, China}
\affiliation{National Astronomical Observatories, Chinese Academy of Sciences, A20 Datun Rd., Chaoyang District, Beijing 100012, China}
\affiliation{Jodrell Bank Centre for Astrophysics, School of Physics and Astronomy, The University of Manchester, Oxford Road, Manchester M13 9PL, UK}

\author{Ismael P\`{e}rez-Fournon}
\affiliation{Instituto de Astrof{\'i}sica de Canarias, C/Vía Láctea, s/n, E-38205 San Cristóbal de La Laguna, Tenerife, Spain}
\affiliation{Universidad de La Laguna, Dpto. Astrof{\'i}sica, E-38206 La Laguna, Tenerife, Spain}

\author{Rui Marques-Chaves}
\affiliation{Instituto de Astrof{\'i}sica de Canarias, C/Vía Láctea, s/n, E-38205 San Cristóbal de La Laguna, Tenerife, Spain}
\affiliation{Universidad de La Laguna, Dpto. Astrof{\'i}sica, E-38206 La Laguna, Tenerife, Spain}

\author{Brice M\`{e}nard}
\affiliation{Department of Physics and Astronomy, Johns Hopkins University, Baltimore, MD 21218, USA}

\begin{abstract}
We present a morphological study of the 17 lensed Lyman-$\alpha$ emitter (LAE) galaxies of the Baryon Oscillation Spectroscopic Survey Emission-Line Lens Survey (BELLS) for the GALaxy-Ly$\alpha$ EmitteR sYstems (BELLS GALLERY) sample. This analysis combines the magnification effect of strong galaxy-galaxy lensing with the high resolution of the \textsl{Hubble Space Telescope} (\textsl{HST}) to achieve a physical resolution of $\sim$80\,pc for this $2<z<3$ LAE sample, allowing a detailed characterization of the LAE rest-frame ultraviolet continuum surface brightness profiles and substructure. We use lens-model reconstructions of the LAEs to identify and model individual clumps, which we subsequently use to constrain the parameters of a generative statistical model of the LAE population. Since the BELLS GALLERY sample is selected primarily on the basis of Lyman-$\alpha$ emission, the LAEs that we study here are likely to be directly comparable to those selected in wide-field narrow-band LAE surveys, in contrast with the lensed LAEs identified in cluster lensing fields. We find an LAE clumpiness fraction of approximately 88\%, significantly higher than found in previous (non-lensing) studies. We find a well-resolved characteristic clump half-light radii of $\sim$350\,pc, a scale comparable to the largest H {\footnotesize \Romannumeral 2} regions seen in the local universe. This statistical characterization of LAE surface-brightness profiles will be incorporated into future lensing analyses using the BELLS GALLERY sample to constrain the incidence of dark-matter substructure in the foreground lensing galaxies.
\end{abstract}


\section{Introduction}

Strong gravitational lensing provides a unique observational tool, both for quantifying the distribution of matter in the lensing objects, and for delivering a magnified view of small and faint sources in the distant universe.
Here we examine the BELLS GALLERY sample of 17 confirmed strong galaxy-galaxy lenses selected spectroscopically from the Baryon Oscillation Spectroscopic Survey (BOSS) of the third Sloan Digital Sky Survey (SDSS-III) and confirmed with high-resolution imaging by the \textsl{Hubble Space Telescope} (\textsl{HST}). The lensing galaxies in the sample are massive red galaxies at redshifts of $z_{\mathrm{lens}} \sim 0.5$, and the sources are Lyman-alpha emitting galaxies (LAEs) at redshifts $2 < z_{\mathrm{LAE}} < 3$. By virtue of their spectroscopic emission-line detection, the lensed LAEs of the BELLS GALLERY sample are comparable to the types of LAEs selected through wide-field narrow-band surveys, but they can be studied in greater detail thanks to the magnifying power of strong lensing.
The primary scientific motivation for the BELLS GALLERY observing program is to use the intrinsically clumpy rest-frame ultraviolet emission in the lensed LAEs as a probe of dark-matter substructure in the lens galaxies. Further discussion of this dark-matter substructure analysis project is provided in \citet{shu2016a, shu2016b}.

This paper presents an analysis of the lensed LAEs of the BELLS GALLERY sample to characterize the clumpiness of their rest-frame UV emission. This information is central to the dark-matter substructure analysis of the sample, so that the surface-brightness perturbations caused by the lensing effects of dark-matter substructure can be detected and characterized statistically. Simultaneously, this paper also provides an unmatched high-resolution characterization of the high-redshift LAE population.

LAE galaxies are defined by a high equivalent width (EW $>$ 20\,\AA) Ly$\alpha$ line and are believed to be composed of extremely large regions of active star formation.  Many efforts have been made to detect and characterize LAE galaxies \citep{elme2009a, elme2009b, bour2007a, cons2003b, cons2004a, ravi2006a,  mand2014a, mood2014a, tacc2010a, gron2011a, guo2015a}.  In general, these galaxies appear as clusters of bright clumps, sometimes with a background of continuum emission.  Evidence suggests that these clumps are larger and brighter than most star forming regions in nearby low redshift galaxies \citep{elme2009a}. Efforts have been made in quantifying mass, star formation rates, gas composition and kinematics, and other LAE properties \citep{nils2009a, tacc2010a, swin2010a, live2015a, hash2017a}.  These have revealed a wealth of information about the early universe, but are ultimately limited by LAE surface brightnesses.  Most studies rely upon stacks of galaxies and can draw only limited inferences about individual LAEs.
The same is true of morphological studies because clump sizes are near the resolution limit of instrumental PSFs (point spread functions) and often cannot be distinguished from point sources \citep{guo2015a}.  As a result, direct imaging studies cannot decisively determine whether the clumps are different in nature from star forming regions in our local universe or if the larger apparent size is merely an artifact of insufficient resolution \citep{tamb2016a}.

Fortunately the magnification due to gravitational lensing improves spatial resolution and allows us to examine scales smaller than the instrumental PSF.
Several studies have used strong gravitational lensing to characterize LAE galaxies and their star-forming regions \citep{jone2010a, swin2010a, swin2012a, barr2014a, live2015a,john2017a}.  These have typical resolutions of $\sim$100\,pc but reach scales as small as 30\,pc, suggesting that clumps are smaller than found in direct imaging studies. 
Here we add to these studies with the single largest sample of lensed galaxies and focus upon a detailed analysis of clump morphology. Because the BELLS GALLERY LAEs are lensed by elliptical galaxies, they will also be subject to fewer systematic modeling biases than the cluster lenses in the surveys above.


The paper is laid out as follows.  In section~\ref{sec:data} we summarize the sample and our source reconstruction approach.  Section~\ref{sec:source_sim} describes the method used to fit surface brightness profiles to individual clumps within the individual galaxies.  In section~\ref{sec:dist_params}, we examine the distributions of the clump parameters, including relative shapes, sizes, surface brightness, and distance from centroid.  We then find a set of analytic probability distribution functions (PDF) to describe these parameters.  In section~\ref{sec:results} we show the results, and finally in section~\ref{sec:discussion} we discuss how our results compare with previous studies and discuss how studies can build upon this work.

\begin{figure*}[t]
\figurenum{1}
  \includegraphics[width=1.0\textwidth]{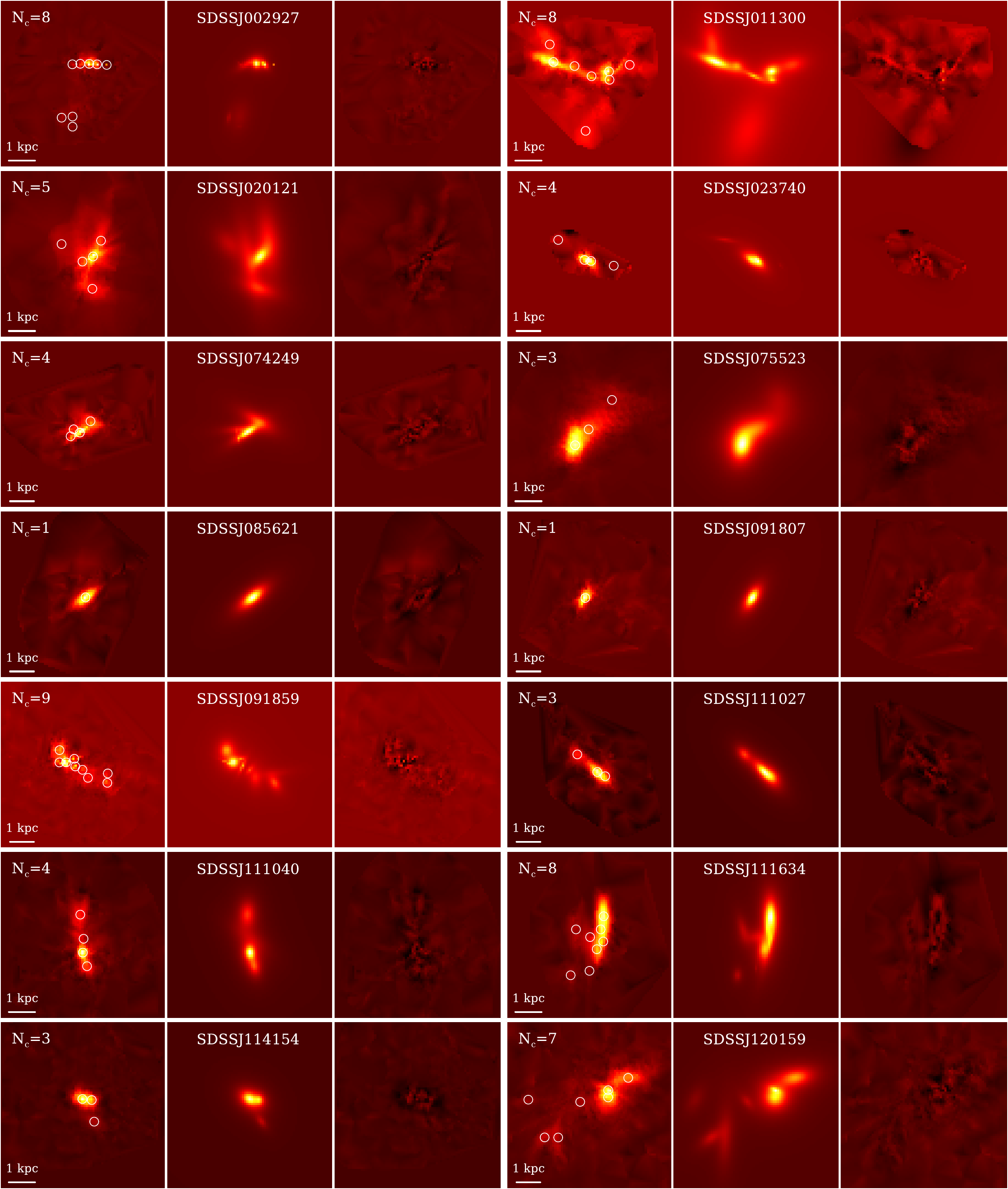}
  \caption{Data (left), model (center), and residuals (right) for all 17 LAE galaxies.  Data is the reconstructed source-plane image derived from pixelized lens modeling.  Lengths scales in the galaxy rest frame are shown.  Intensity scales vary between galaxies.  The numbers of identified clumps are indicated for each galaxy and clump locations are marked with white circles.  The model residuals are correlated due to our image reconstruction method.}
\end{figure*}
\begin{figure*}[t]
  \includegraphics[width=1.0\textwidth]{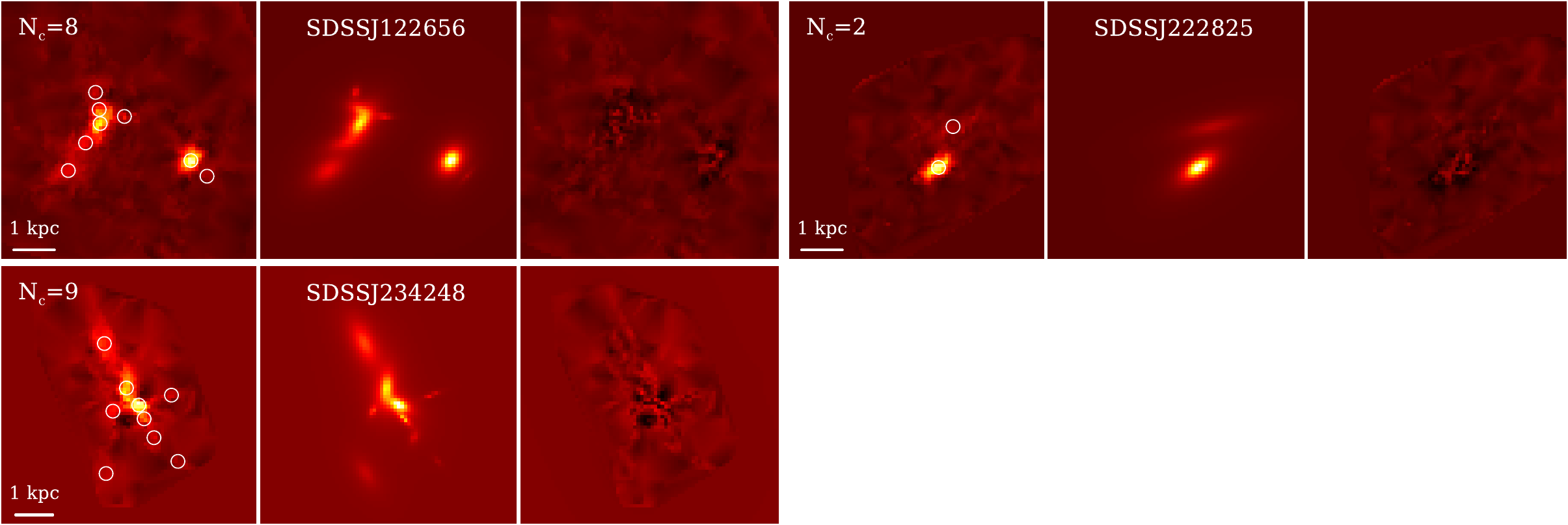}
  \caption{Continued.}
  \label{mosaic}
\end{figure*}

\section{Data}
\label{sec:data}

Candidate lensed LAE galaxies were selected from the SDSS-III/BOSS spectroscopic data set \citep{daws2013a}. BOSS is a cosmological redshift survey designed primarily to constrain the properties of dark energy using the baryon acoustic oscillation (BAO) feature imprinted in the large-scale structure traced by galaxies. By virtue of the large number of spectra ($\sim10^6$) obtained by BOSS, rare objects such as the lenses of the BELLS GALLERY sample can also be found in significant numbers. The lens systems that we analyze here were identified through the presence of prominent emission lines that were not consistent with emission from the target galaxy or other low-redshift interloper objects, as detailed in \citet{shu2016a}.

Follow-up imaging of the best candidates was conducted with the Wide Field Camera 3 (WFC3) aboard \textsl{HST} under Cycle 23 GO Program 14189.  In total, 21 targets were observed for one orbit each through the F606W filter ($\sim$1500\,\AA--1900\,\AA for the source galaxies).  Image inspection and analysis \citep{shu2016b} shows that 17 of the candidates are grade A (definite) gravitational lenses.

We use reconstructed source images found by \citet{shu2016b}.  \citet{shu2016b} used a pixelized lens modeling procedure to constrain the parameters of a smooth lens mass model and to reconstruct high-resolution images of the source LAEs. The models used a variation on the ``semi-linear'' framework of \citet{warr2003a} (see also \citealt{koop2006a,vege2009a,nigh2015a}), which does a linear inversion for the pixelized source-plane image along with a non-linear optimization of the lens mass model parameters.  To prevent the model from over-fitting noise, we included a regularization parameter.  This regularization parameter was tuned until the $\chi^2$ of the pixelized model is similar to that of the $\chi^2$ found by modeling the source as several S\`{e}rsic clumps.
The effect of varying the regularization parameter is discussed in section~\ref{sec:results}.

Because lens modeling accounts for the PSF of \textsl{HST}, the reconstructed LAE images are effectively deconvolved, and are taken directly as input data for the analysis in this paper.  These images can be seen in the left panel of each group in Figure~\ref{mosaic}.  The effective angular resolution is on the order of 0.01 arcsec, corresponding to a typical spatial resolution of $\sim$80\,pc.  This is an order of magnitude better than direct observations with \textsl{HST},
and is comparable to the $\sim$100\,pc resolution found in most other gravitational lensing studies.
Similar resolutions may be achieved with the next generation Extremely Large Telescopes.

\section{Modeling}

Our overall goal is to use the data to constrain the parameters of a generative statistical model of the rest-frame UV continuum surface-brightness characteristics of the $2<z<3$ LAE population.  In particular, we want to:
\begin{enumerate}
\item Characterize the physical sizes, arrangements, and multiplicities of the clumps that make up the rest-frame ultraviolet emission of $2 < z < 3$ LAEs.
\item Simulate realistic high-resolution optical images of $2 < z < 3$ LAEs.
\item Determine probability density functions (PDFs) for multiple clump LAE surface-brightness models that can be applied within a hierarchical Bayesian analysis of evidence for dark-matter substructure in the BELLS GALLERY sample.
\end{enumerate}

We proceed in two main stages: first, to detect and parametrically model the individual clumps in each reconstructed image (\S\ref{sec:source_sim}); and second, to model the statistical distribution of these clump parameters across the sample (\S\ref{sec:dist_params}).

\subsection{Source Fitting}
\label{sec:source_sim}

We model each LAE source galaxy in our sample as a collection of clumps. This procedure involves individual clump detection, adoption of a parametric profile for modeling individual clumps, and simultaneous optimization of the model parameters for all clumps in each LAE.

Our initial clump detection is done manually through visual inspection. We also investigated an automated clump finder like that described in \citet{guo2015a}, but found that this approach had difficulty resolving overlapping clumps that were more easily distinguished by visual inspection.  Spatial noise correlations also gave rise to false clumps which were better rejected visually.

The steps that we use for clump detection and modeling are as follows:
\begin{enumerate}
\item  Set a heuristic brightness detection threshold in each LAE image at 10\% of the maximum image pixel brightness.  This threshold is used to reject spurious clump identifications associated with correlated noise peaks, especially in the outlying regions of the image.

\item Inspect a 3D surface contour map in combination with a 2D color image of each LAE to select pixels that represent clump centers.  Individual clumps must be separated by at least two pixels.

\item Using our adopted clump surface-brightness profile (see below), set initial guesses for clump model parameters.  For the radius parameter, we set $r_c$ to be the clump full width at half max (FWHM) divided by 2.  For the central surface-brightness parameter ($i_c$), we take the count value in the central pixel.  We estimate the axis ratio ($q_c$) and position angle (PA) parameters by eye.  Each parameter is allowed to vary independently between clumps.

\item Manually tune the guesses to decrease the model residuals, using a reduced chi-squared ($\chi^2_r$) as the metric. Pixel errors are derived from the lens-model reconstruction and depend both upon Poisson noise in the \textsl{HST} images and lens model parameters (see \citet{shu2016b} for more details).  Note that this $\chi^2_r$ is not strictly correct due to correlated noise from the lens-modeling reconstruction, but it is a good indicator of relative fit quality.

\item When manual adjustments have no significant impact, use the \texttt{lmfit} non-linear fitting routine \citep{newv2014a} for a final optimization of parameters. To prevent runaway parameters, we constrain the clump centroid parameters $x_c$, $y_c$ to be within $\pm0.5$\,pixels 
and constrain the clump radius and central surface-brightness parameters to be at most a factor of 3 from their original estimates. The position angles are not bounded, and the axis ratios are initially allowed to vary between $0.01 < q_c < 100$.  For a best-fit $q_c>1$ we apply the transformation $q_c' = 1/q_c$ and $PA' = PA + \frac{\pi}{2}$ so the final $q_c$ ranges between 0.01 and 1. In some cases, unphysically flattened axis ratio fits are obtained. In these cases, we re-fit the data with the same initial guesses, but convert $r_c$ and $q_c$ into major and minor axis lengths ($a_c$ and $b_c$) that are allowed to increase or decrease by a factor of 3. This procedure eliminates the unphysical axis ratios.    \label{item:fit}


\item Repeat the fitting procedure of step~\ref{item:fit}, updating starting parameters and parameter bounds according to the results of the previous step. Continue to repeat until $\chi^2_r$ no longer improves. By iterating on parameters and parameter bounds in this way, we achieve a converged model without allowing the clump models to diverge entirely from their initially detected identities.
\end{enumerate}







\begin{figure*}[t]
\gridline{\fig{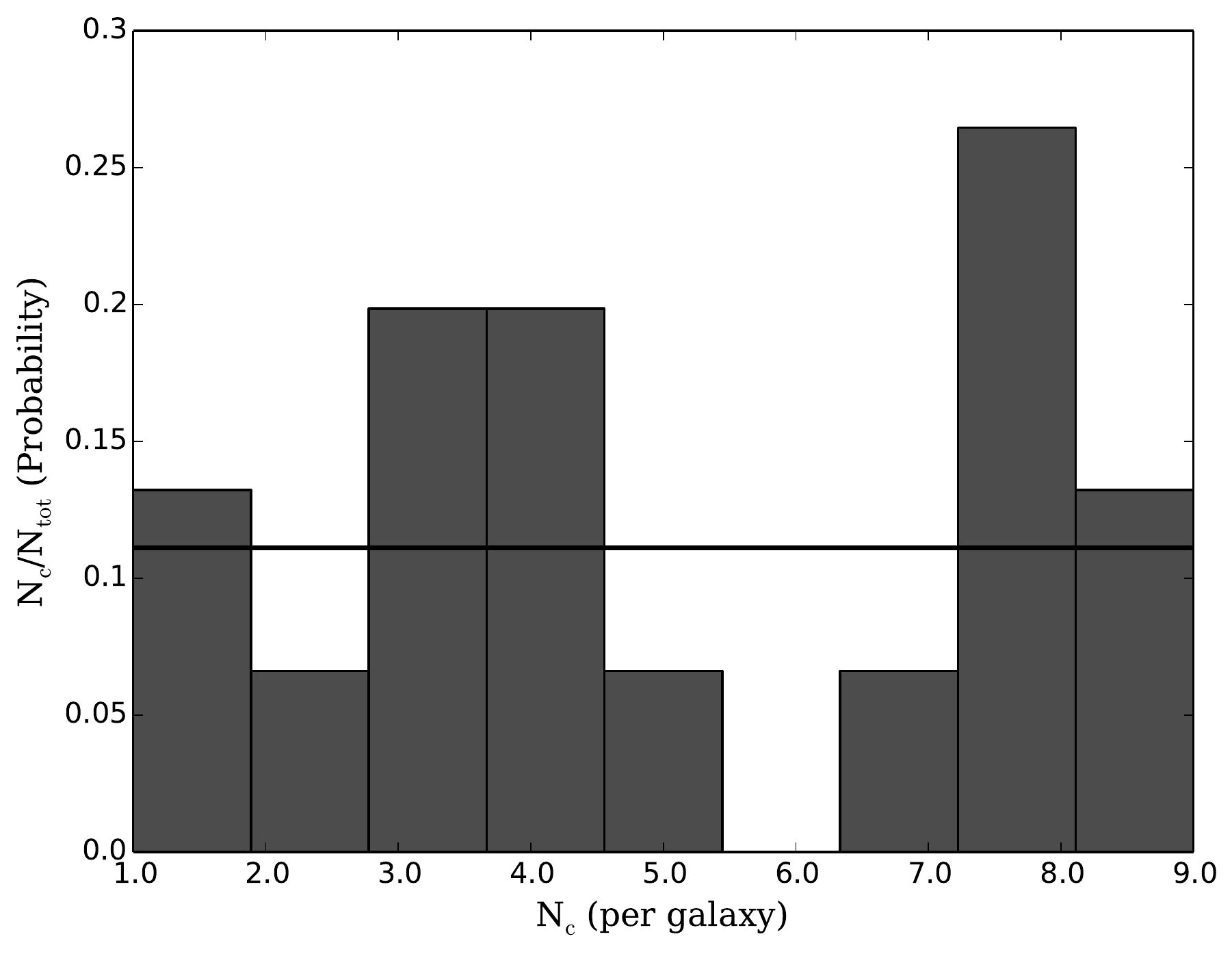}{0.5\textwidth}{}
          \fig{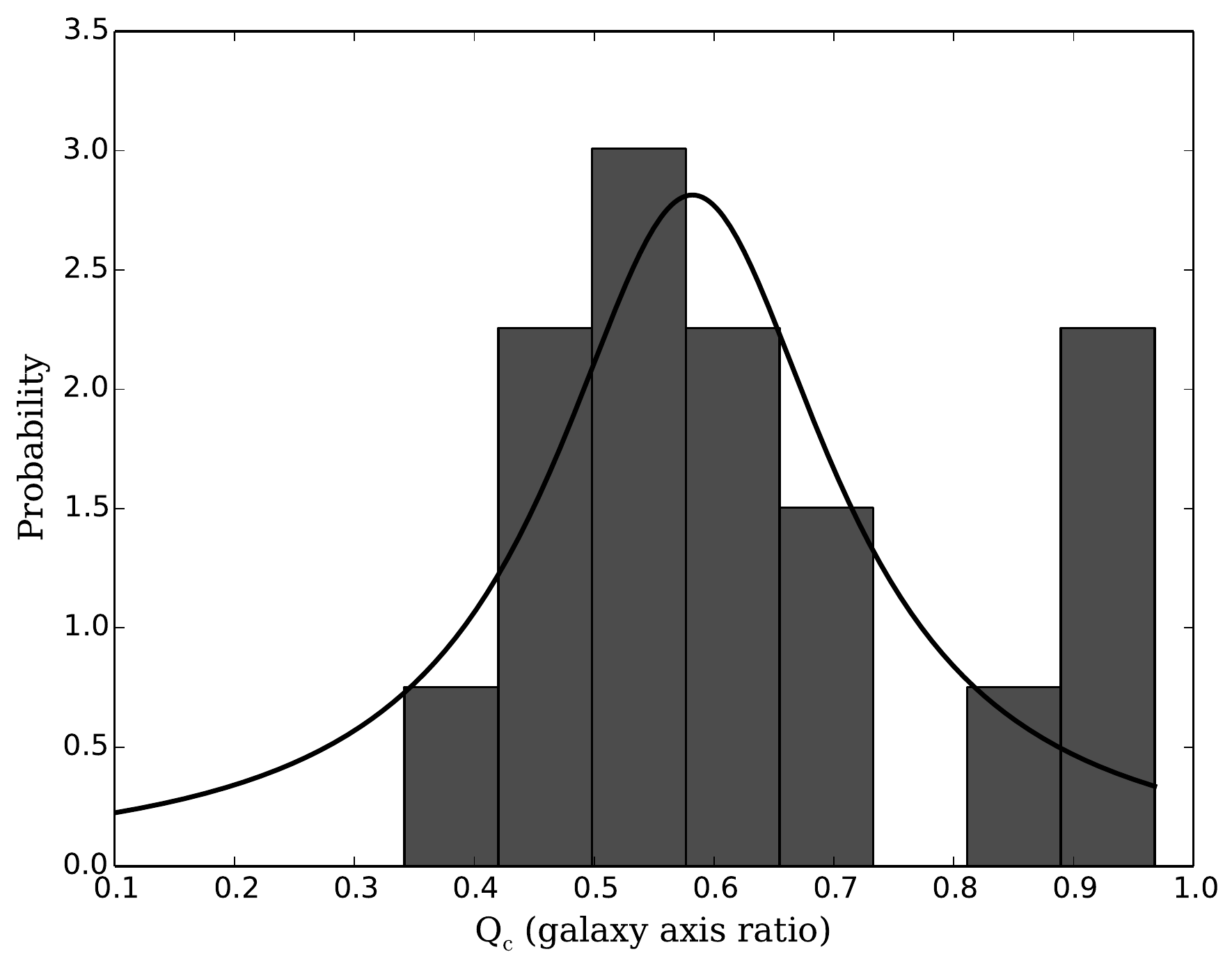}{0.5\textwidth}{}}
\caption{\textit{Left:} Distribution of the number of clumps found in the 17 galaxies.  No strong trend is observed, thus a uniform distribution is assumed in the subsequent analysis.  \textit{Right:} Mean quadrupole moment for each galaxy.  This gives an estimate of the galaxy axis ratio, which appears centrally peaked.  Each plot is overlaid with a black line indicating the best-fit PDF for the chosen distribution (see section~\ref{sec:dist_params}).}
\label{fig:nclumps_gal}
\end{figure*}

Our chosen surface-brightness profile is the Elson, Fall, and Freeman (EFF) profile \citep{elso1987a, schw2004a, ryon2015a},

\begin{equation}
\label{eff_eqn}
I(r) = i_c(1+r^2/r_c^2)^{-\eta}
\end{equation}

\noindent
used to fit resolved H {\footnotesize \Romannumeral 2} regions in local galaxies.  We generalize $r$ to an elliptical radius by

\begin{equation}
\label{r_ell}
r = \sqrt{q_c(x-x_c)^2 + (y-y_c)^2/q_c}
\end{equation}

\noindent
where $q_c$ is the axis ratio and $x_c$ and $y_c$ are the horizontal and vertical clump centroids respectively.  The scale factor $r_c$ can be converted to the half-light radius by

\begin{equation}
\label{eff_half_light}
r_{1/2} = r_c\sqrt{(1/2)^{\frac{1}{1-\eta}}-1}
\end{equation}

\noindent
We also examined the S\`{e}rsic profile \citep{sers1968a, ciot1999a} which is ubiquitous in galaxy fitting routines.  However, we report results only for the EFF profile with the exponent fixed at $\eta=3/2$, a typical value found in the local universe \citep{ryon2015a}.


\begin{table}
\centering
\begin{tabular*}{0.4685\textwidth}{ |c|l| }
\hline
\textbf{Parameter} & \textbf{Description} \\ \hline
$Q_g$ & mean LAE axis ratio/quadrupole moment \\ \hline
$N_c$ & number of clumps (per galaxy) \\ \hline
$x_c$ & horizontal clump centroid \\ \hline
$y_c$ & vertical clump centroid \\ \hline
$d_c$ & clump radial distance from centroid \\ \hline
$r_c$ & clump radius \\ \hline
$i_c$ & peak clump surface brightness \\ \hline
$q_c$ & clump axis ratio \\ \hline
$PA_c$ & clump position angle \\ \hline
\end{tabular*}
\caption{Description of clump parameters for reference.}
\label{table:param_names}
\end{table}

The raw data, the final EFF model, and the residuals are shown in Figure~\ref{mosaic} for each galaxy.  Physical length scales are roughly the same for each image, but the intensity (color) scale varies slightly between each galaxy.  The mottled appearance of the residuals is a result of correlated noise inherent in gravitationally lensed source reconstruction (section~\ref{sec:data}).

During this procedure, we did not include a component for a diffuse background continuum.  This component is often included in other surveys, but many of these encompass multiple bands \citep{guo2015a}.  \citet{elme2009a} found that if a continuum was evident it was often only at wavelengths redder than our rest frame UV data.  Moreover any diffuse background emission associated with a lensed LAE will be partly absorbed into the smooth model of the foreground lens galaxy and effectively subtracted.  Thus it is not surprising that our data show no evidence of such a continuum.

Clump radii and relative positions are converted to parsecs using the LAE redshifts from \citet{shu2016a} and a fiducial cosmological model with $\Omega_m$ = 0.274, $\Omega_{\Lambda}$ = 0.726, and $H_0$ = 70\,kms$^{-1}$\,Mpc$^{-1}$ (WMAP7, \citealt{koma2011a}).

\subsection{Parameter Distribution}
\label{sec:dist_params}

\begin{figure*}[t]
\gridline{\fig{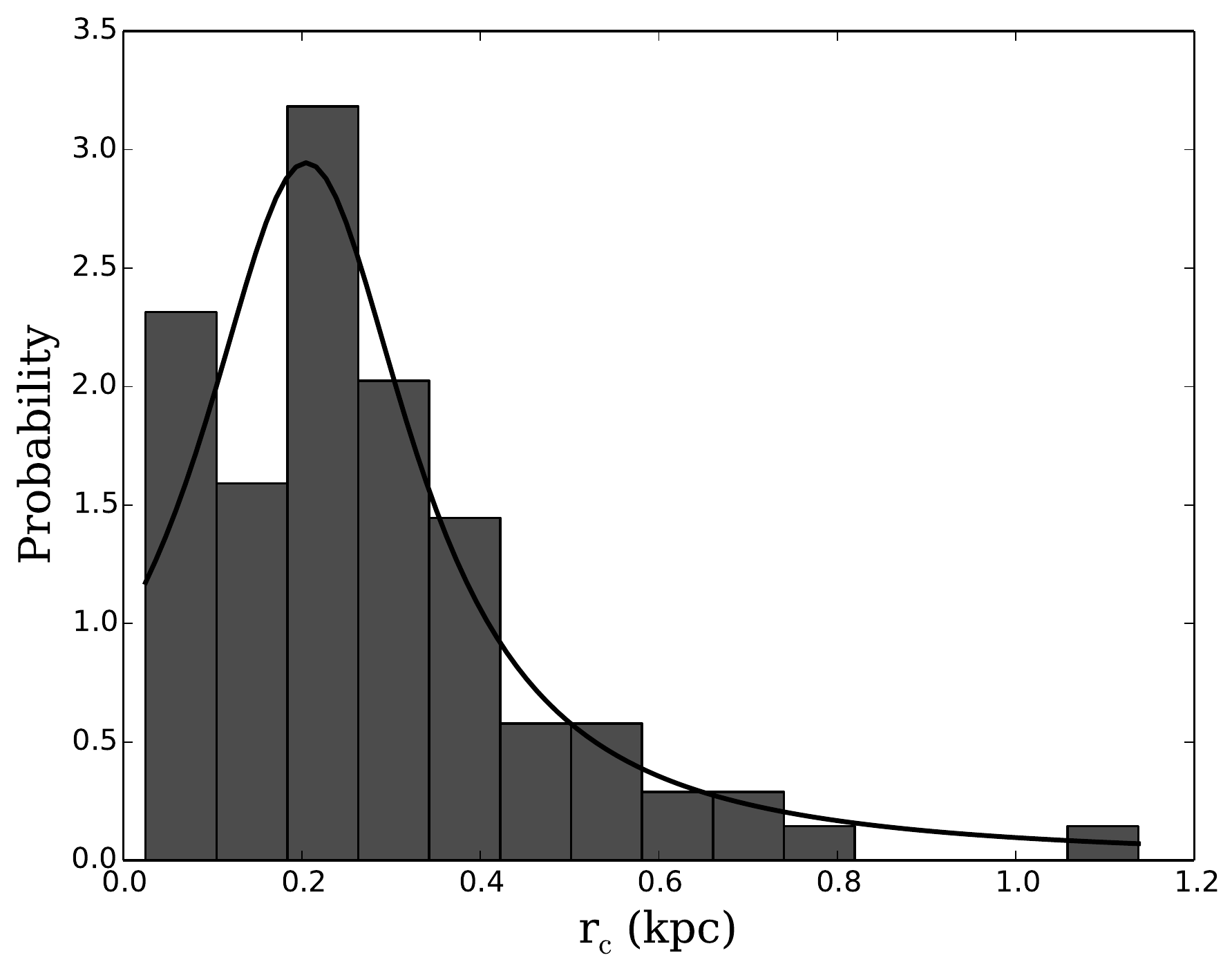}{0.5\textwidth}{}
          \fig{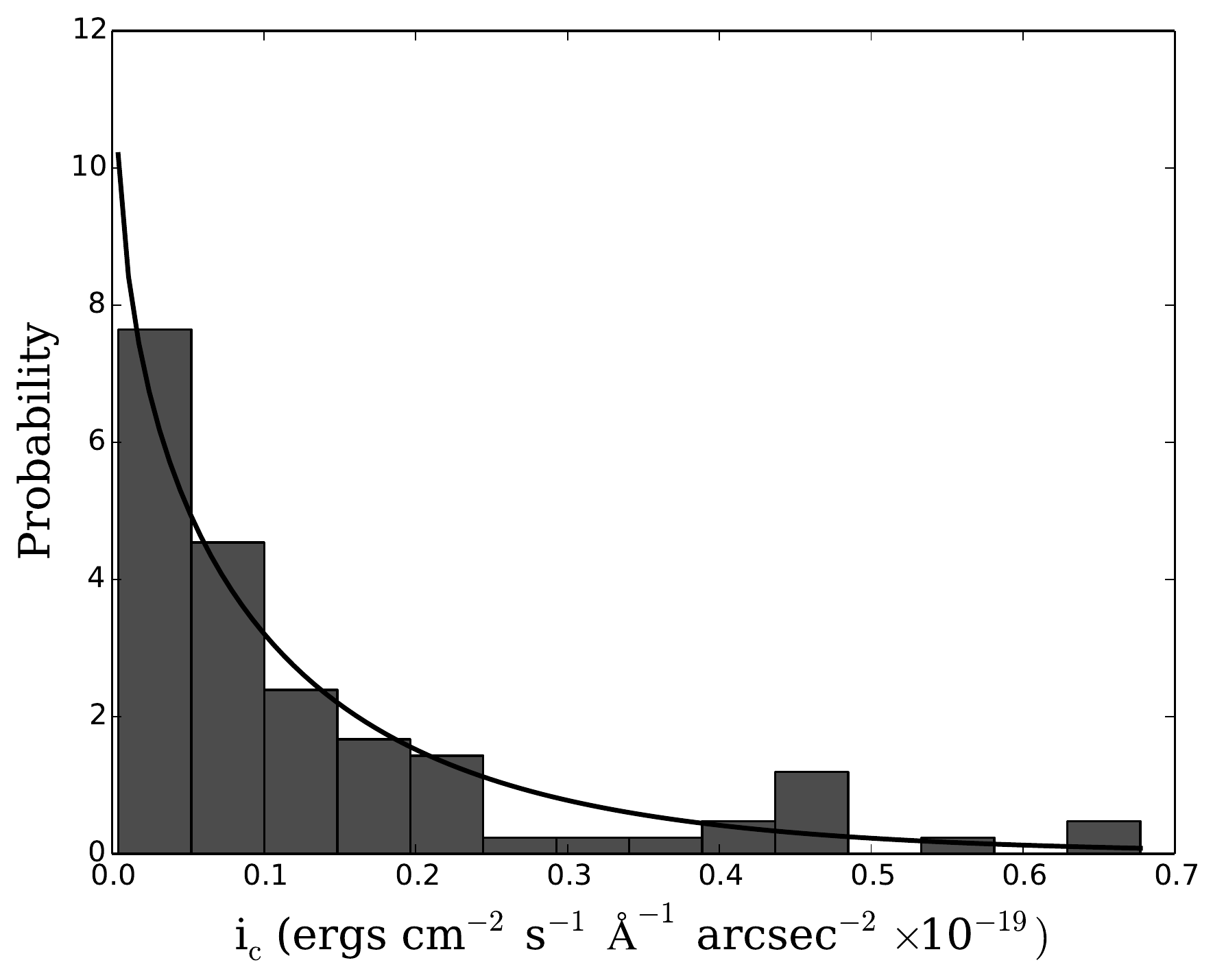}{0.5\textwidth}{}}
\gridline{\fig{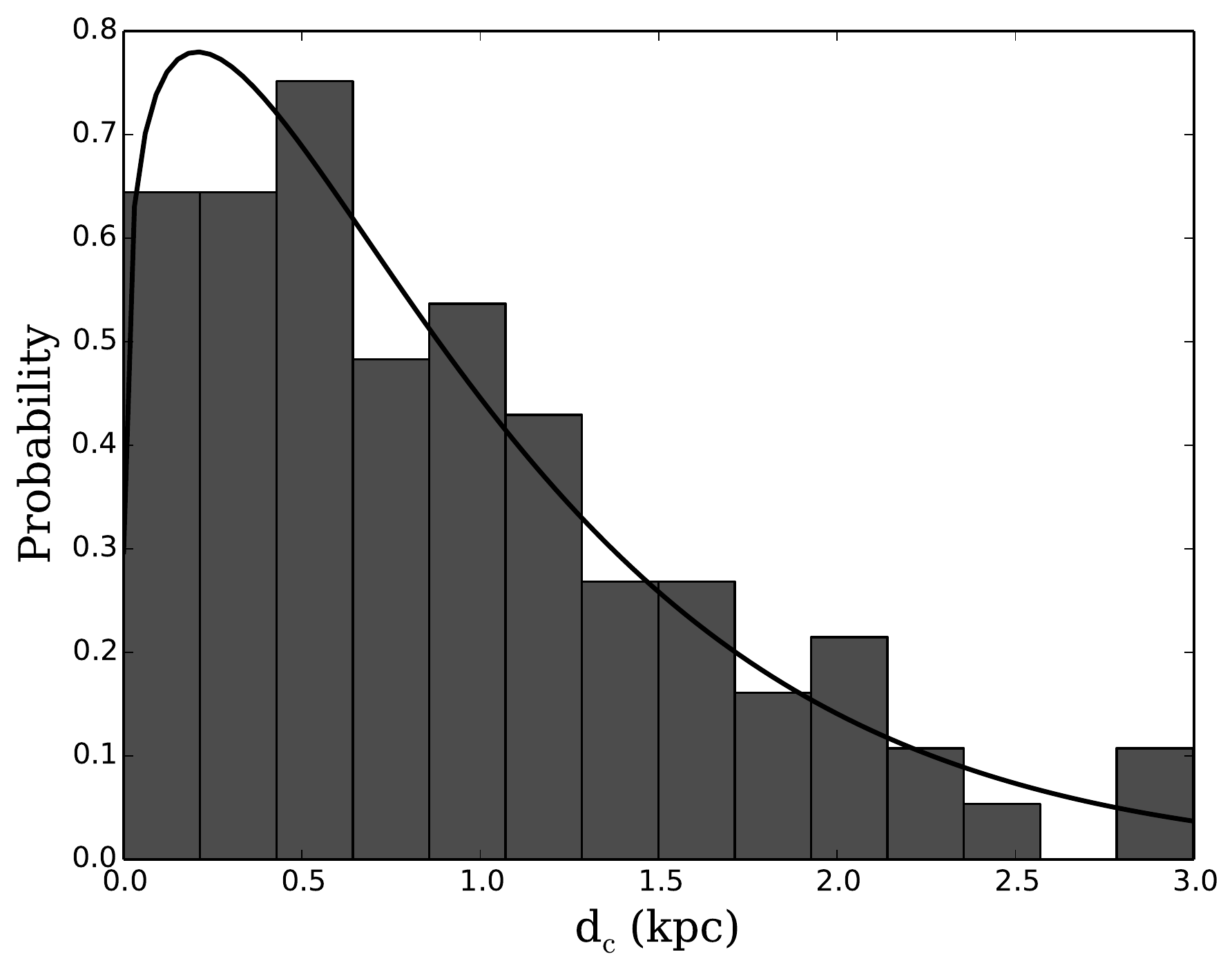}{0.5\textwidth}{}
          \fig{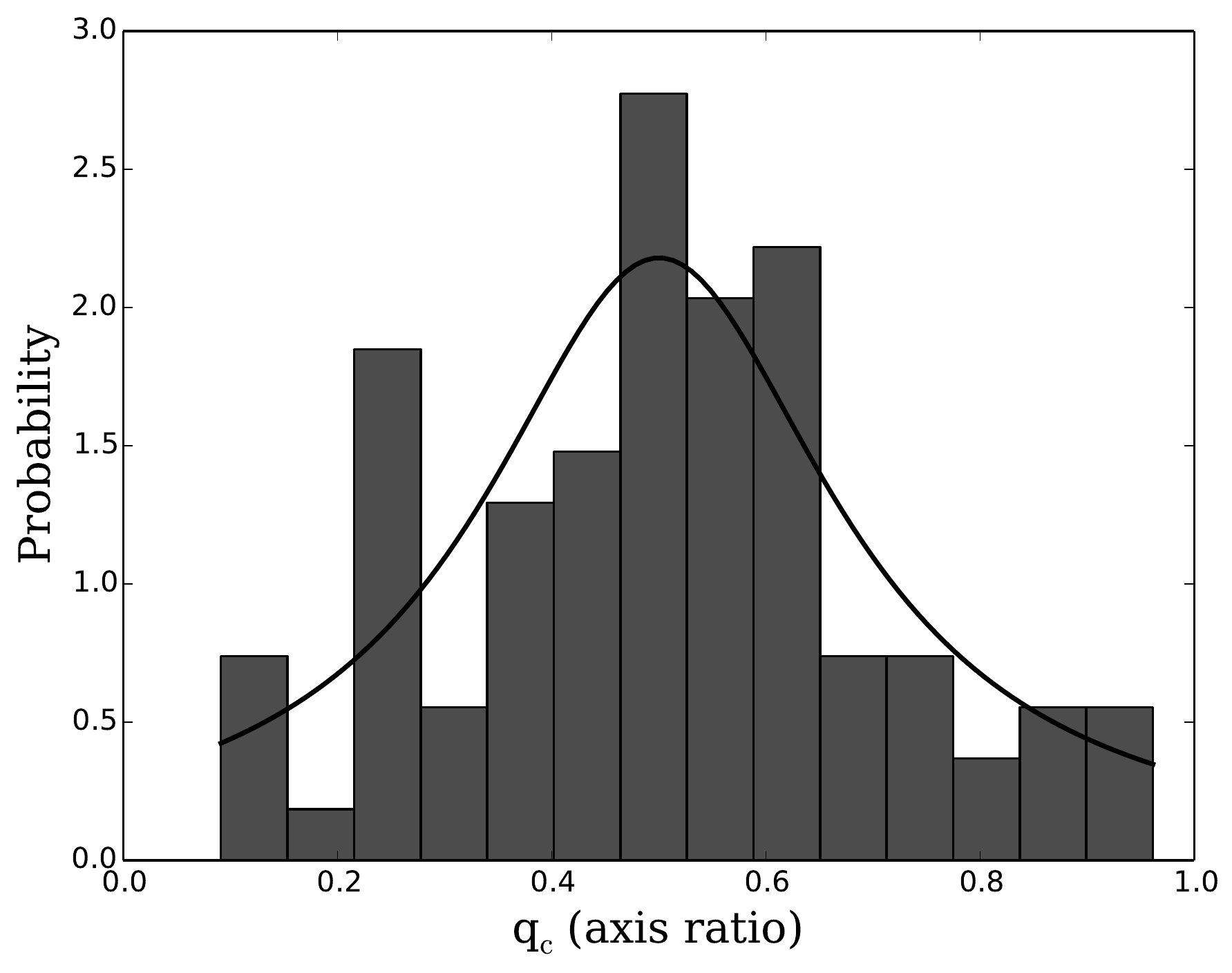}{0.5\textwidth}{}}
  \caption{Histograms of the clump parameters.  \textit{Top left:} The scale radius of each clump, $r_c$.  \textit{Top right:} The peak surface brightness, $i_c$.  \textit{Bottom left:} The distance of the clump center from the light-weighted centroid, $d_c$.  \textit{Bottom right:} The clump axis ratio, $q_c$.  The best fit distributions are overlaid in solid black.}
  \label{model_param_hist}
\end{figure*}

Given the models of the 17 LAEs, we now examine the statistical distribution of clump parameters from Table~\ref{table:param_names} across the sample.
The two parameters, the number of clumps $N_c$ and the mean surface brightness quadrupole moment $Q_g$ are measured for each LAE. The distributions for these two parameters are shown in Figure~\ref{fig:nclumps_gal}.

The 87 clumps across the sample are characterized by six parameters: the position $x_c, y_c,$ size $r_c,$ peak brightness $i_c,$ axis ratio $q_c,$ and position angle $PA_c$.  For the purposes of statistical analysis, we combine $x_c$ and $y_c$ into a single parameter $d_c$, the radial distance from the galaxy light-weighted centroid, which characterizes the radial distribution of clumps within an LAE.  
The average ellipticity of the distribution of clumps within an LAE is captured by the quadrupole parameter $Q_g$.
The distribution of clump position angles shows no preference for either radial or tangential alignment with respect to the overall LAE geometry, and as such we assume these position angles to be randomly distributed.
The distributions of the four remaining clump parameters ($d_c, r_c, i_c,$ and $q_c$) are shown in Figure~\ref{model_param_hist}.

\begin{figure*}
  \includegraphics[width=\textwidth]{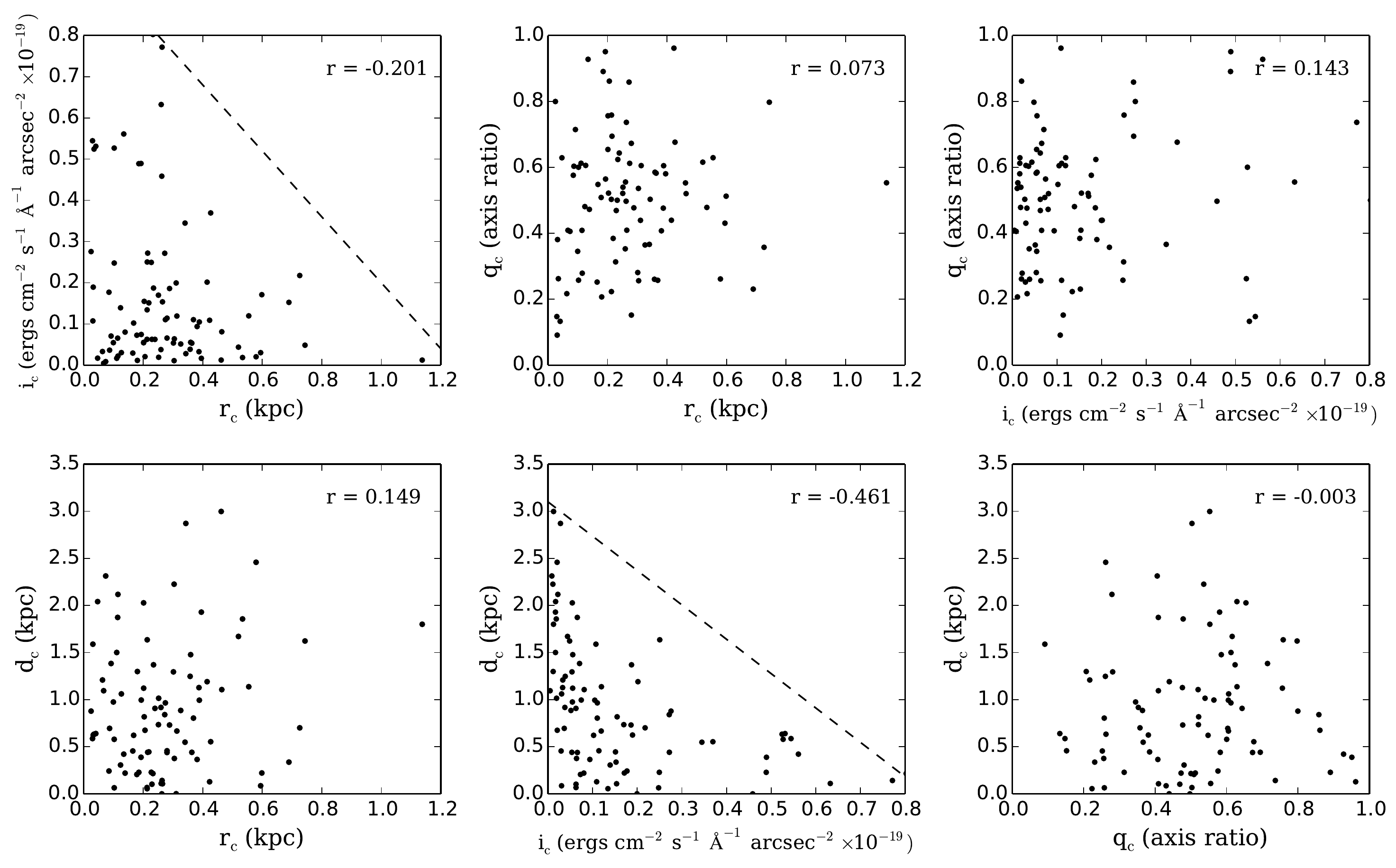}
  \caption{Scatterplots of parameters with their associated correlation coefficients.  Correlations are identified between $i_c$ and $r_c$ (top left) and between $d_c$ and $i_c$ (bottom center).  To re-create these correlations, we impose a linear cutoff (dashed lines) above which random draws are rejected in later simulations.}
  \label{model_param_scatter}
\end{figure*}

Before modeling the distributions of these parameters across the sample, we apply completeness corrections, surface-brightness corrections, and K-corrections

To estimate incompleteness, we first measure the noise level in the 17 input images.  Though correlated, this noise is still approximately Gaussian.  From this, we identified the mean 1$\sigma$ noise level as 0.005 photon counts/second/pixel, with 40\% scatter.  We then randomly select our model clumps from the sample of 87 and compare their central-pixel brightness to the noise level.  This brightness is a function of half-light radius and peak surface brightness and preferentially selects against large radius, low peak surface brightness clumps.  Those with central-pixel brightness below the 1$\sigma$ noise level are considered undetected.  We repeat this process until each clump is well sampled, at least 10,000 times.  We then weight the radius and surface brightness of each point by the inverse of its detection rate.  In practice, this effect is small; only three clumps show a need for incompleteness corrections, all with weights below a factor of 5.  Even if the noise threshold is increased, the completeness corrections remain small.

For the surface-brightness corrections, we account for the impact of the variable source redshift by correcting all galaxies to a common reference redshift of $z_{ref}=2.6$.   This effect is included by multiplying each per-{\AA}ngstr\"{o}m surface brightness value by a factor of $(1+z_{gal})^5/(1+z_{ref})^5$.
Because we lack sufficient spectral information to empirically determine K-corrections, we adopt the corrections in \citet{vand2010a} (Figure 7 top) from a qualitatively similar selection of galaxies.
To convert these to our sample we note that the i-band corrections in the paper between redshifts 3-4 correspond to a central UV wavelength of 1900\,\AA--1500\,\AA, which matches the central wavelength of the F606W filter between redshifts $2.2<z<3$.
The corrections are approximately linear over this redshift range and
can be approximated by

\begin{equation}
K = -0.055(z-2.6)\,\mathrm{mag}
\label{eqn:kcorr}
\end{equation}

\noindent
for a correction to $z_{ref}=2.6$.


Before we select models for the probability density functions (PDFs) of the clump parameters, we consider the possibility of correlations between various parameters, as shown in Figure~\ref{model_param_scatter}
We determine the significance of these correlations by calculating Pearson's correlation coefficient for each of the parameter pairs using 1000 bootstrap samplings from the 87 identified clumps.
We find a significant correlation, greater than 2$\sigma$ (2.5th and 97.5th percentile), between $d_c$ and $i_c$ and between $r_c$ and $i_c$.  We attempted to model these parameters with joint distributions, but were unable to adequately re-create the observed correlation.  Instead, we model each clump parameter as uncorrelated and impose a linear cutoff above which randomly drawn clumps are rejected (see dashed lines in Figure~\ref{model_param_scatter}).

\begin{table*}
\centering
\def\arraystretch{1.5}
\begin{tabular*}{0.8595\textwidth}{ |l|l|p{1.125in}|p{1.125in}|p{1.125in}| } 
\hline
\textbf{Quantity} & \textbf{Distribution} & \textbf{Parameters \qquad \qquad (standard reg.)} & \textbf{Parameters \qquad \qquad  (low reg.)} & \textbf{Parameters \qquad \qquad (high reg.)} \\ \hline
\multirow{2}{1in}{$Q_{g}$ (galaxy axis ratio)} & \multirow{2}{1in}{Cauchy (eqn.~\ref{cauchy_eqn})} & $Q_o \ = \ \ 0.58^{+.41}_{-.04}$ & $Q_o \ = \ \ 0.67^{+.31}_{-.13}$ & $Q_o \ = \ \ 0.62^{+.10}_{-.03}$ \\
\cline{3-5} & & $\gamma_Q \ = \ \ 0.14^{+.20}_{-.06}$ & $\gamma_Q \ = \ \ 0.33^{+.24}_{-.12}$ & $\gamma_Q \ = \ \ 0.11^{+.07}_{-.05}$ \\  \hline 
\multirow{2}{1in}{$q_{c}$ (clump axis ratio)} & \multirow{2}{1in}{Cauchy (eqn.~\ref{cauchy_eqn})} & $q_o \ \ = \ \ 0.50^{+.03}_{-.03}$ & $q_o \ \ = \ \ 0.42^{+.45}_{-.32}$ & $q_o \ \ = \ \ 0.53^{+.07}_{-.06}$ \\
\cline{3-5} & & $\gamma_q \ \ = \ \ 0.20^{+.05}_{-.04}$ & $\gamma_q \ \ = \ \ 0.44^{+.42}_{-.44}$ & $\gamma_q \ \ = \ \ 0.27^{+.13}_{-.08}$ \\ \hline 
\multirow{2}{1in}{$r_{c}$ (radius, kpc)} & \multirow{2}{1in}{Cauchy (eqn.~\ref{cauchy_eqn})} & $r_o \ \ = \ \ 0.20^{+.03}_{-.04}$ & $r_o \ \ = \ \ 0.18^{+.02}_{-.03}$ & $r_o \ \ = \ \ 0.33^{+.04}_{-.04}$ \\
\cline{3-5} & & $\gamma_r \ \ = \ \ 0.15^{+.04}_{-.03}$ & $\gamma_r \ \ = \ \ 0.11^{+.02}_{-.02}$ & $\gamma_r \ \ = \ \ 0.18^{+.04}_{-.03}$ \\ \hline 
\multirow{2}{1in}{$i_{c}$ (counts)} & \multirow{2}{1in}{Weibull (eqn.~\ref{weibull_eqn})} & $\lambda_i \ \ = \ \ 0.13^{+.02}_{-.02}$  & $\lambda_i \ \ = \ \ 0.17^{+.02}_{-.02}$ & $\lambda_i \ \ = \ \ 0.08^{+.01}_{-.01}$ \\ \cline{3-5} 
 & & $k_i \ \ = \ \ 0.87^{+.09}_{-.11}$ & $k_i \ \ = \ \ 1.02^{+.07}_{-.06}$ & $k_i \ \ = \ \ 0.83^{+.11}_{-.10}$ \\ \hline 
\multirow{2}{1in}{$d_{c}$ (distance, kpc)} & \multirow{2}{1in}{Weibull (eqn.~\ref{weibull_eqn})} & $\lambda_d \ \ = \ \ 1.00^{+.10}_{-.09}$ & $\lambda_d \ \ = \ \ 0.94^{+.09}_{-.09}$ & $\lambda_d \ \ = \ \ 0.87^{+.15}_{-.13}$ \\ \cline{3-5} 
 & & $k_d \ \ = \ \ 1.18^{+.12}_{-.10}$ & $k_d \ \ = \ \ 1.22^{+.10}_{-.09}$ & $k_d \ \ = \ \ 1.17^{+.12}_{-.22}$ \\ \hline 
\end{tabular*}
\caption{Probability distributions for the clump parameters.  Column 3 gives best-fit PDF parameters for the quantities $Q_g$, $r_c$, $i_c$, $d_c$, and $q_c$ given the standard distributions in Figure~\ref{model_param_hist} while columns 4 and 5 show results using reconstructions with lower and higher regularization parameters.}
\label{pdf_table}
\end{table*}

There is no a priori distribution expected for any of our clump-model parameters, and therefore we explored a number of standard distributions (Gaussian, Weibull, Cauchy) to find a mathematical model that is sufficiently but not overly flexible.
We adopt a Cauchy distribution

\begin{equation} \label{cauchy_eqn}
P(x) = \Big[\pi\gamma\Big(1 + \Big(\frac{x-x_{0}}{\gamma}\Big)^{2}\Big)\Big]^{-1}
\end{equation}

\noindent
for the galaxy axis-ratio parameter $Q_g$, clump axis-ratio parameter $q_c$, and the clump radius $r_c$ and a Weibull distribution

\begin{equation} \label{weibull_eqn}
P(x) = \frac{k}{\lambda}\Big(\frac{x}{\lambda}\Big)^{k-1}e^{-(x/\lambda)^k}
\end{equation}

\noindent
for the distance from light-weighted centroid $d_c$ and the clump surface brightness $i_c$. The Cauchy PDFs are truncated at x=0 and above the maximum value found in section~\ref{sec:source_sim}

For each of these PDFs, we determine the best-fit parameters by maximizing the log-likelihood function

\begin{equation} \label{log_likelihood_eqn}
\ln L(\boldsymbol{\theta}|\mathbf{x}) = \sum_{i=1}^{n_{c}}\ln P(\mathbf{x_i},\boldsymbol{\theta})
\end{equation}

\noindent
where P is the PDF model (eqn~\ref{cauchy_eqn} or~\ref{weibull_eqn}), the $\mathbf{x}$ are the measured clump parameters and the $\boldsymbol{\theta}$ are the PDF parameters.





Finally, we quantify the effect of the choice of lens-modeling regularization parameter (see section~\ref{sec:data}) on our PDF model results. We repeat the analysis detailed in sections~\ref{sec:source_sim} and~\ref{sec:dist_params} for two alternative sets of lensing-reconstructed source images: one generated with a regularization parameter decreased by a factor of 10 relative to the standard value, and the other with a regularization parameter increased by a factor of 10.  The impact of this parameter is discussed in the following section.

\section{Results}
\label{sec:results}

Table~\ref{pdf_table} summarizes the results, and the model distributions are shown in black overlaying the histograms in Figures~\ref{fig:nclumps_gal} and~\ref{model_param_hist}.  Parameter uncertainties are determined using bootstrap resampling with 1000 iterations and we report the 16th and 84th percentiles, corresponding to $1\sigma$ in the Gaussian limit.  PDF parameter estimates for the two alternative regularization scenarios are also shown.

\begin{figure*}[t]
  \includegraphics[width=1.0\textwidth]{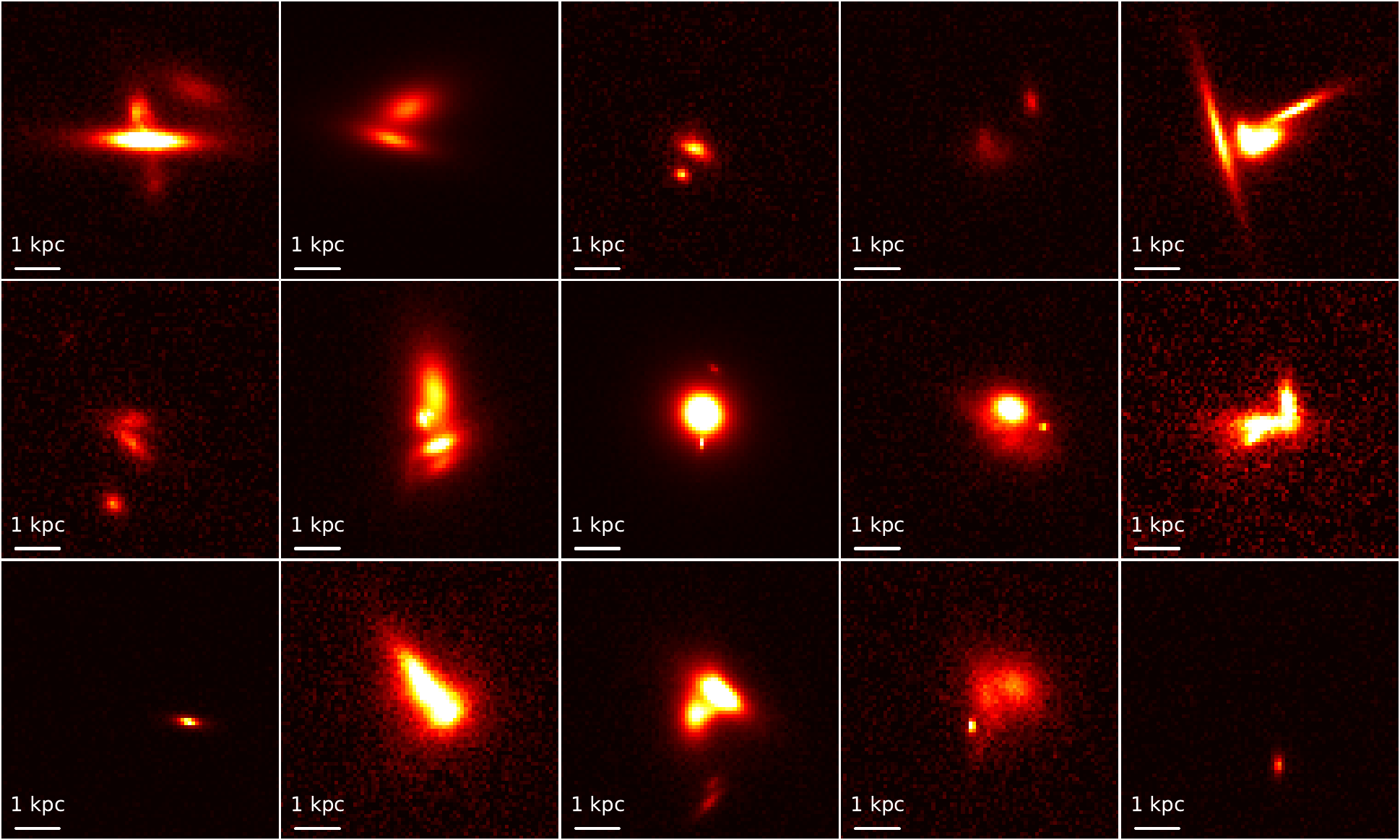}
  \caption{Simulated z=2.6 galaxies drawn from fitted PDFs and shown with identical intensity (color) scales.  These resemble the observed LAEs shown in Figure~\ref{mosaic}, confirming that the fitted PDFs describe the source galaxy morphology.}
  \label{sim_gxy_mosaic}
\end{figure*}

We find the following results from our analysis:

\begin{enumerate}

\item Out of our 87 total clumps, we find 1--9 clumps per galaxy, with an approximately uniform distribution in number and 88\% of galaxies having more than one clump ($f_{clumpy}=88\%$).  Lowering the regularization parameter, which reduces the smoothing, gave 88 total clumps with 1--8 clumps per galaxy and $f_{clumpy}=94\%$.  Higher regularization, which increases the smoothness, decreases the total clump count to 61 with 1--7 clumps per galaxy, yet finds the same $f_{clumpy}=94\%$.

\item The characteristic half-light clump radius is 350\,pc
with 85\% having half-light radii above 160\,pc (2 pixels) suggesting that most clumps are fully resolved in our images.  A lower regularization parameter has no significant impact on the characteristic radius, but a higher regularization parameter increases it to 630\,pc.

\item The distances of the clumps from the light-weighted centroid is relatively small, with a typical scale length of 1\,kpc, and with most clump centers contained within 2\,kpc of the galactic center.  This result is robust against model choice and regularization parameter. This distance is comparable to the typical galaxy half-light radius, which ranges from 0.4-1.6\,kpc with a median at 0.8\,kpc.

\item Clump axis ratios peak around 0.5 and galaxy axis ratios peak around 0.6.  This peak is roughly constant with regularization, but the width of the clump distribution broadens for both alternative regularization choices, particularly with low regularization where the distribution becomes more consistent with a uniform distribution than a Cauchy.


\end{enumerate}

In calculating these results, we examined the blurring impact of observational noise. We first generate simulated galaxy images (see Figure~\ref{sim_gxy_mosaic}), with parameters drawn from the fitted distributions, and add typical Gaussian background noise.  We then fit the images to derive output clump parameters and fit the resulting distributions.  The input and output clump parameters generally agree to within 1$\sigma$ and the output model PDF parameters agree with the inputs to 2$\sigma$ or better.  We thus conclude that observational noise has a negligible impact on the parameter estimates.

\section{Discussion and Conclusions}
\label{sec:discussion}

Although the BELLS GALLERY sample is relatively small compared to LAEs selected from wide-field, narrow-band surveys, enhanced spatial resolution from combining gravitational lensing and \textsl{HST} imaging allows more detailed surface-brightness characterization relative to other samples of LAEs at the same redshifts.

We found 1--9 clumps per galaxy, which is higher than most direct imaging studies.  While \citet{elme2005a} found as many as 10 clumps, \citet{guo2015a} found a strong preference for fewer clumps and a lower $f_{clumpy}$ of $\sim55\%-60\%$.  A larger number of clumps are, in part, expected due to our increased resolution.  Indeed, we identify a strong correlation between number of clumps and magnification (from \citet{shu2016b}, Table 2).  In lensed images of SDSS J1110+6459 at a resolution of 30\,pc \citet{john2017a,rigb2017a} identified over 20 clumps with radii between 30 and 50\,pc.   
If we restrict the accounting to clumps contributing more than 8\% of the total UV luminosity, as proposed in \citet{guo2015a}, we find that the total number of clumps drops to 65 with a range of 1--7 clumps per galaxy, although $f_{clumpy}$ remains at 88\%.

Our typical clump radius of 350\,pc is comparable to the average clump radius of $\sim$320\,pc found in cluster-lensing studies \citep{live2015a}, but significantly lower than the 750-900\,pc found in direct imaging studies \citep{bond2012a}.  Most LAE clumps are not fully resolved without a combination of strong lensing and high-resolution imaging. Even in the BELLS GALLERY sample, we expect some clumps to be unresolved, especially considering the 30--50\,pc scale clumps identified in \citet{john2017a}. Nevertheless, at our typical effective spatial resolution of 80\,pc we appear to resolve a characteristic peak in the clump radius distribution. We conclude that high-redshift LAE clumps are significantly larger than most local H {\footnotesize \Romannumeral 2} regions, having typical radii similar to the largest giant H {\footnotesize \Romannumeral 2} regions detected in the local universe \citep{fuen2000a,monr2007a,wisn2012a}.

Our measurements of overall LAE galaxy sizes confirm that these galaxies are spatially small, at least in the UV spectral range, with size scales on the order of 0.4-1.6\,kpc.  This size scale is consistent with direct imaging surveys that find LAE half light radii between 1-1.5\,kpc \citep{bond2012a, malh2012a}.  The spatial distribution of the clumps, proportional to P($d_c$)/$d_c$, is consistent with an exponential function with a scale radius of $\sim$0.3\,kpc.

The LAE galaxy ellipticities in our sample are similar, but slightly smaller than those measured in other studies.  \citet{gron2011a} found that ellipticities peak near 0.55 in redshift z=3.1 LAEs and \citet{ravi2006a} show a similar peak of $\sim$0.55 for z$\sim$3 Lyman break galaxies (LBGs).  These studies also found a skew toward high ellipticities (low axis ratios) which we did not find here, but our sample is too small to consider this difference significant.

The details of the BELLS GALLERY selection function are different from those of other LAE surveys, and in particular vary between galaxies due to different magnification factors.  However, the fact that these lenses are selected based on Ly$\alpha$ emission suggests that our sample should encompass a qualitatively similar set of galaxies to those found in wide-field, ground-based, narrow-band surveys. In comparison, lensed LAEs selected in galaxy \textit{cluster} fields are generally identified based on H$\alpha$ or H$\beta$ lines, making it likely that our LAE sample is a closer analog to field LAE samples.

In addition to providing a quantitative characterization of the $2<z<3$ LAE population, the statistical model of LAE clumpiness that we present here can be used to simulate mock LAEs, like those in Figure~\ref{sim_gxy_mosaic}, for other studies, and to provide prior probabilities on LAE surface-brightness structure for our forthcoming analysis of dark-matter substructure in the foreground lensing galaxies.


\acknowledgments

Support for program \# 14189 was provided by NASA
through a grant from the Space Telescope Science Insti-
tute, which is operated by the Association of Universities
for Research in Astronomy, Inc., under NASA contract
NAS 5-26555.

Funding for SDSS-III was provided by the Alfred P.
Sloan Foundation, the Participating Institutions, the Na-
tional Science Foundation, and the U.S. Department
of Energy Office of Science. The SDSS-III website is
http://www.sdss3.org/.

SDSS-III was managed by the Astrophysical Research
Consortium for the Participating Institutions of the
SDSS-III Collaboration including the University of Ari-
zona, the Brazilian Participation Group, Brookhaven
National Laboratory, Carnegie Mellon University, Uni-
versity of Florida, the French Participation Group,
the German Participation Group, Harvard University,
the Instituto de Astrof{\'i}sica de Canarias, the Michigan
State/Notre Dame/JINA Participation Group, Johns
Hopkins University, Lawrence Berkeley National Labora-
tory, Max Planck Institute for Astrophysics, Max Planck
Institute for Extraterrestrial Physics, New Mexico State
University, New York University, Ohio State University,
Pennsylvania State University, University of Portsmouth,
Princeton University, the Spanish Participation Group,
University of Tokyo, University of Utah, Vanderbilt Uni-
versity, University of Virginia, University of Washington,
and Yale University.

This work was supported by the National Natural Science Foundation of China (Grant No. 11333003, 11390372 to SM, and Grant No. 11603032 to Y.S.).

CSK is supported by NSF grant ASF-1515876

IPF and RMC acknowledge support from the Spanish Ministerio de Economia y Competitividad (MINECO) under grant number ESP2015-65597-C4-4-R

\facilities{\textsl{HST} (WFC3), Sloan}

\software{Anaconda, \texttt{lmfit} \citep{newv2014a}}




\clearpage

\bibliography{bibtex_archive}
\bibliographystyle{aasjournal}

\end{document}